\documentclass[10pt]{article}
\usepackage{amssymb,amsmath}
\usepackage{graphicx}
\usepackage{color}
\usepackage{hyperref}
\usepackage{curves,multicol}
\usepackage{geometry}
\usepackage{times}
\usepackage{bm}
\usepackage{ulem}
\usepackage{ucs}
\usepackage[utf8x]{inputenc}
\usepackage[intoc]{nomencl}
\usepackage{float}
\usepackage{subfig}
\usepackage{units}
\usepackage{titlesec}
\usepackage{curves,multicol}
\usepackage{mathptmx}
\usepackage{lscape}
\usepackage{nopageno}
\usepackage[round]{natbib}
\usepackage{xcolor}
\usepackage[labelfont=bf]{caption}
\usepackage{lipsum}   
\usepackage[none]{hyphenat} 
\def\dashL{\bm{\mbox{--~--~--}}}
\def\dotL{\bm{\mbox{$\cdot\ \cdot\ \cdot$}}}

\def\Lbox{\mbox{---}~{\hspace*{-.1in}\tiny $\square$}\hspace*{-.1in}~\mbox{---}}
\def\Lcirc{\mbox{---}~{\hspace*{-.12in}$\circ$}\hspace*{-.12in}~\mbox{---}}

\makeatletter

\renewcommand{\section}{\@startsection
{section}%                   % the name
{1}%                         % the level
{0mm}%                       % the indent
{-\baselineskip}%            % the before skip
{0.5\baselineskip}%          % the after skip
{\normalfont\bfseries\MakeUppercase}} % the style

\renewcommand{\subsection}{\@startsection
{subsection}%                   % the name
{2}%                         % the level
{0mm}%                       % the indent
{0.5\baselineskip}%            % the before skip
{0.25\baselineskip}%          % the after skip
{\bfseries\normalsize}} % the style

\makeatother

\begin{document}
\sloppy

\setcounter{secnumdepth}{-1} 

\vspace*{-2cm}
% \begin{flushright} \vbox{
% 35th Symposium on Naval Hydrodynamics\\
% Nantes, France, 8 -- 12 July, 2024}
% \end{flushright}

%\vskip0.65cm
\vskip1.3cm
\begin{center}
\textbf{\LARGE
Data Assimilation-based Simultaneous Phase-Resolved Ocean Wave and Ship Motion Forecast\\[0.35cm]
}

\Large G. Wang and Y. Pan\\ 

% \Large G. Wang$^{2}$, Y. Pan$^{1}$, and R. Roe$^{2}$\\ 

(Department of Naval Architecture and Marine Engineering, University of Michigan, Ann Arbor, MI USA)\\
\vspace*{0.25cm}

\end{center}

\begin{multicols*}{2}

\section{Abstract}
% The phase-resolved ocean wave field information and ship responses are crucial for the overall performance of maritime activities. However, due to the imperfect knowledge

% and the chaotic nature of the nonlinear wave equations

This paper presents a data-assimilation (DA)-based approach to forecast the phase-resolved  wave evolution process and ship motion, which is developed by coupling the high-order spectral method (HOS), ensemble Kalman filter (EnKF), and a Cummins-equation-based ship model (CMI). With the developed EnKF-HOS-CMI method, the observation data for wave, ship, or both can be incorporated into the model, therefore producing the optimal analysis results. The developed method is validated and tested based on a synthetic problem on the motions of an irregular wave field and a box-shaped free-floating ship. We show that the EnKF-HOS-CMI method achieves much higher accuracy in the long-term simulation of nonlinear phase-resolved wave field and ship motion in comparison with the HOS-CMI method. Also, the ship parameters are estimated accurately by using a parameter-augmented state space in EnKF.

% can estimate both wave field and ship states, by sequentially assimilating observed wave, ship, or both data.

%. With the development of the remote sensing and computational technologies, it is now possible to reconstruct the initial phase-resolved ocean surface from radar measurements and launch a nonlinear wave model such as the high-order spectral (HOS) method to predict the wave evolution in real time. However, due to the unavoidable errors in model configurations (e.g., initial conditions and physical parameters) and the chaotic nature of the nonlinear wave equations, the prediction by HOS can deviate quickly from the true dynamics. Recent studies, including those of the authors, have shown that this dilemma can be eased to some extent by incorporating the wave observational data into models via data assimilation methods such as ensemble Kalman filter (EnKF). In this work, we aim at further improvement of wave prediction accuracy and realizing the simultaneous floating structure motion forecast. This is realized by coupling HOS, EnKF, and a Cummins-equation-based floating structure model, and including the observed floating structure motion as one additional data source. Through numerical testing, it is shown that the new integrated approach not only provides accurate floating structure motion forecast, but also uplifts the wave prediction accuracy compared to the state-of-the-art single-source (wave) data methods.

\section{Introduction}
\label{SECintroduction}

The prediction of wave-induced ship responses (e.g., motions, strain, and surface pressure) are crucial for the overall safety and efficiency of maritime activities. They enable better health monitoring and decision-making. In the meanwhile, it is equally meaningful to reconstruct the encountered wave profiles based on the observed ship responses, which can serve as initial conditions to predict the future (even short-term) wave field and ship motions. In this regard, two types of methods to estimate the {\it{in-situ}} wave information from the measured ship responses have been developed: one for wave spectra, and the other for the encountered incident wave sequences.

The methods for estimation of wave spectra from structural responses have been well developed~\citep{pascoal2005ocean,nielsen2006estimations,tannuri2003estimating,nielsen2020ocean, chen2020estimation}. Formulated in the frequency domain, these methods utilize frequency analyses (e.g., Fast Fourier Transform, FFT) and predetermined transfer functions to estimate the onsite one-dimensional or directional wave spectra. However, it should be noted that although the estimated wave spectrum parameters can aid the maritime operations to some extent, it is impossible to realize the deterministic prediction of the future ship-encountered wave field and ship responses with this type of methods. For example, the potential ship capsize induced by extreme waves cannot be captured, which can however lead to the significant loss of wealth and lives. 

In contrast, the methods to estimate the encountered incident wave sequences, which are formulated in the time domain, have the capability to realize the deterministic prediction for both wave field and ship motions, and therefore are receiving more and more attentions recently.~\cite{koterayama2002study} applied FFT analysis on the observed motions of a wave buoy to determine the phase information for each component of the incident wave field. However, the accuracy of this method is sensitive to the time length of the observed buoy motion, and short-time series (e.g.~$<10~\text{min}$) may lead to significant deviation from the reference.~\cite{takami2022application,takami2023estimation} developed a prolate spheroidal wave functions (PSWF)-based approach, which can determine the incident wave profiles from short-time sequences of ship motion data (i.e. $2-5\text{min}$). However, it should be noted that these methods based on linear analysis ignores the nonlinear components of the problem, which may lead to deteriorated performance when the latter becomes important.
%While it should be noted that, when predicting the future wave field and ship responses based on the reconstructed wave sequences from the PSWF-based approach, the results may quickly deviate from the truth, due to the linear assumption in its derivation process. 

In this work, we develop a data assimilation (DA)-based approach to realize the phase-resolved estimation of non-linearly evolving wave fields from the ship motions. This is realized by extending the author's previous work on DA-based non-linear phase-resolved ocean wave reconstruction and forecast framework, namely the EnKF-HOS method by coupling an ensemble Kalman filter and the high-order spectral (HOS) method~\citep{wang2021phase,wang2022phase}. Through extensive test cases, it has been demonstrated that the EnKF-HOS method can realize the high-accuracy phase-resolved ocean wave reconstruction and forecast by sequentially assimilating observed wave data. Here we add to the EnKF-HOS method a Cummins-equation-based ship motion model (CMI)~\citep{ogilvie1964recent}. By doing this, we build the correlation between the ship motions and encountered wave field, which provides the foundation for the wave field reconstruction. More importantly, if the assimilation result is used as initial condition for forecast, the method has the capability to more accurately capture the future wave evolution and ship motion. Finally, in comparison with EnKF-HOS method, the newly developed EnKF-HOS-CMI can work with different types of data, i.e. ship motion data, wave data, or both, therefore it has favorable robustness in different application scenarios.

The paper is organized as follows. We first introduce the mathematical formulation and methodologies used in this study, including the EnKF-HOS-CMI coupled framework, as well as its each component, HOS, EnKF, and Cummins equation. The validity of the proposed algorithm is tested through numerical experiments on a wave-ship coupled simulation, considering an irregular wave field and a box-shaped free-floating ship, with noisy initial conditions and imperfect ship parameters. We show that, with sufficient data assimilation, the forecast errors for both the wave field and ship motion are significantly reduced compared to the solutions without DA. In addition, the simultaneous estimation of ship parameters are realized. We conclude with some discussions on the implementation of this method for realistic wave-ship coupled problems.

\section{Mathematical formulation and methodology}
\label{sec:PF}

\subsection{Problem statement}
\noindent We consider the motions of an irregular wave field and a box-shaped free-floating ship (e.g., a barge), with the (noisy) wave data, ship motion data, or both available. We denote the surface elevation and surface potential, reconstructed from the wave data, as $\eta_{\text{m},j}(\boldsymbol{x})$ and $\psi_{\text{m},j}(\boldsymbol{x})$, and ship motion data as $\boldsymbol{S}_{\text{m},j}$, with $j=0,1,2,3, \cdots$ the index of time $t$ and $\boldsymbol{x}$ the spatial coordinates. It is also assumed that the error statistics associated with $\eta_{\text{m},j}(\boldsymbol{x})$, $\psi_{\text{m},j} (\boldsymbol{x})$, and $\boldsymbol{S}_{\text{m},j}$ is known {\it a priori} from the inherent properties of the measurement equipments (e.g., wave buoy and motion reference unit). 

In addition to the measurements, we have available a wave model that is able to simulate the evolution of the ocean surface (in particular $\eta(\boldsymbol{x},t)$ and $\psi(\boldsymbol{x},t)$), and a ship model that is able to simulate the ship motion $\boldsymbol{S}(t)$ in the wave field. Our purpose is to incorporate wave measurements $(\eta_{\text{m},j}(\boldsymbol{x}), \psi_{\text{m},j}(\boldsymbol{x}))$, ship motion measurements $\boldsymbol{S}_{\text{m},j}$, or both into the model simulation sequentially in an optimal way such that the analysis of the states $\eta_{\text{a},j}(\boldsymbol{x})$, $\psi_{\text{a},j}(\boldsymbol{x})$, and $\boldsymbol{S}_{\text{a},j}$ are most accurate.   

\subsection{The general EnKF-HOS-CMI coupled framework}
\label{sec:enkfhos}
\begin{figure}[H]
\includegraphics[width=\linewidth]{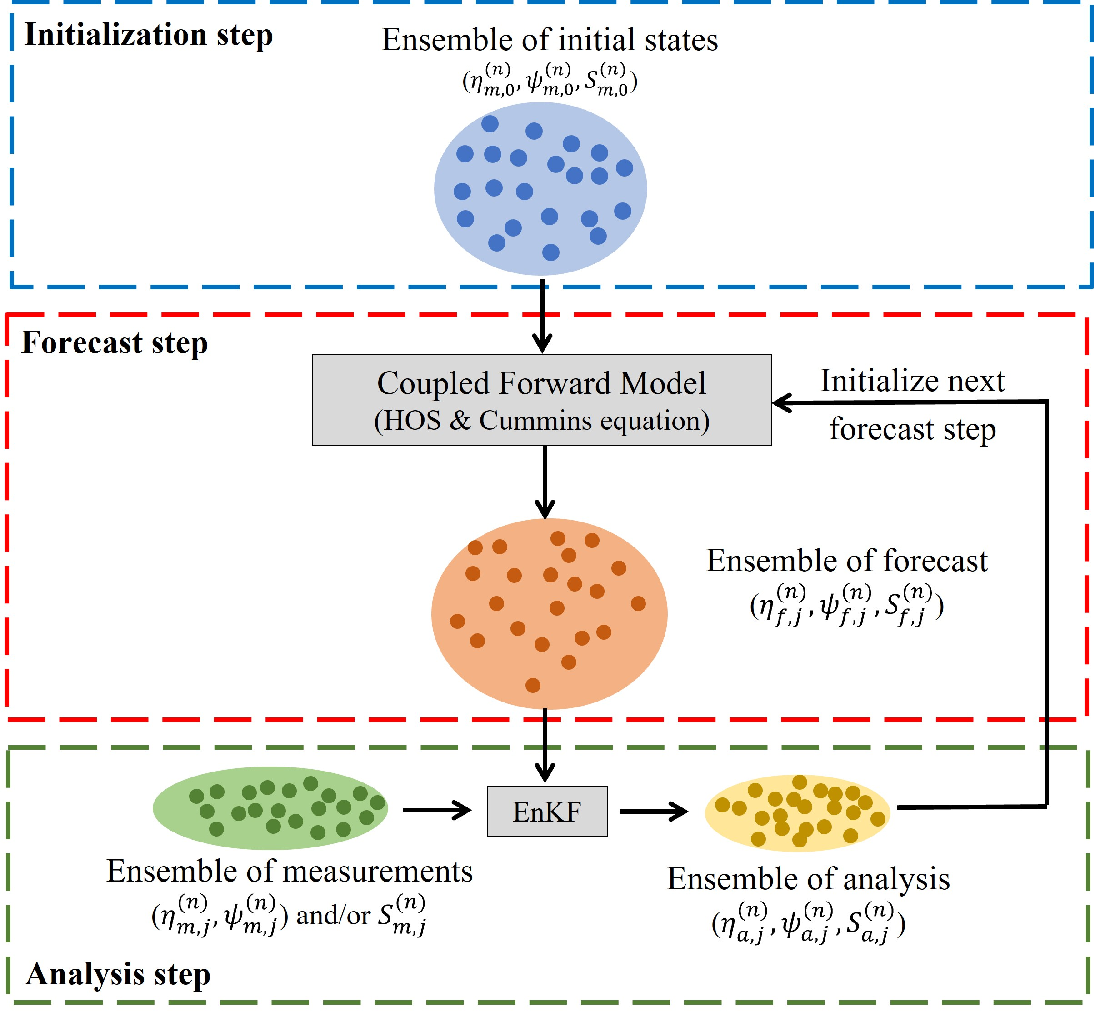}
\caption{Schematic illustration of the EnKF-HOS-CMI coupled framework. The size of ellipse represents the amount of uncertainty.} \label{fig:scheme}
\end{figure}
In this study, we use HOS and Cummins equation to simulate the nonlinear phase-resolved wave evolution process and ship motion, respectively, coupled with EnKF for DA. Figure~\ref{fig:scheme} shows a schematic illustration of the proposed EnKF-HOS-CMI coupled framework. At initial time $t=t_0$, measurements $(\eta_{\text{m},0}, \psi_{\text{m},0}, \boldsymbol{S}_{m,0})$ are available, according to which we generate ensembles of perturbed measurements, $(\eta_{\text{m},0}^{(n)}, \psi_{\text{m},0}^{(n)}, \boldsymbol{S}_{\text{m},0}^{(n)} )$, $n=1,2,...,N$, with $N$ the ensemble size, following the known measurement error statistics (i.e., sampling from a given distribution). A forecast step is then performed, in which an ensemble of $N$ HOS and CMI simulations are conducted, taking $(\eta_{\text{m},0}^{(n)}, \psi_{\text{m},0}^{(n)}, \boldsymbol{S}_{\text{m},0}^{(n)})$ as initial conditions for each ensemble member $n$, until $t=t_1$ when the next measurements become available. At $t=t_1$, an analysis step is performed where the model forecasts $(\eta_{\text{f},1}^{(n)}, \psi_{\text{f},1}^{(n)}, \boldsymbol{S}_{\text{f},1}^{(n)})$ are combined with new measurements $(\eta_{\text{m},1}^{(n)}, \psi_{\text{m},1}^{(n)})$, $ \boldsymbol{S}_{\text{m},1}^{(n)}$, or both to generate the analysis results $(\eta_{\text{a},1}^{(n)}, \psi_{\text{a},1}^{(n)}, \boldsymbol{S}_{\text{a},1}^{(n)}$). A new ensemble of HOS and ship motion simulations are then performed taking $(\eta_{\text{a},1}^{(n)}, \psi_{\text{a},1}^{(n)}, \boldsymbol{S}_{\text{a},1}^{(n)}$) as initial conditions, and the procedures are repeated for $t=t_2, t_3, \cdots$ until the desired forecast time $t_{\text{max}}$ is reached. 

We next describe in detail the key components in the EnKF-HOS-CMI method, the HOS method, Cummins equation, and the EnKF method. Regarding the generation of measurement ensembles from given error statistics, we refer the readers to the references \cite{wang2021phase} and \cite{wang2022phase} for details. For simplicity, in the following description we assume that the gravitational acceleration and fluid density are unity (so that they do not appear in equations) by choices of proper time and mass units.

\subsection{HOS method and Cummins equation}
\label{sec:hos}
\noindent Here we assume that the ship dimensions are much smaller than the peak wave length, and therefore the impacts of the floating ship on the wave field are negligible. Given the initial wave fields $(\eta_{\text{m},0}^{(n)}, \psi_{\text{m},0}^{(n)})$  or $(\eta_{\text{a},j}^{(n)}, \psi_{\text{a},j}^{(n)})$ with $j \geq 1$, for each ensemble member $n$, the evolution of the wave state from $t_j$ to $t_{j+1}$ is solved by integrating a set of nonlinear wave equations in Zakharov form:

\begin{eqnarray}
    &&\frac{\partial\eta(\boldsymbol{x},t)}{\partial t} + \frac{\partial\psi(\boldsymbol{x},t)}{\partial \boldsymbol{x}} \cdot \frac{\partial\eta(\boldsymbol{x},t)}{\partial \boldsymbol{x}}\nonumber \\ 
    &&-\left[1+\frac{\partial\eta(\boldsymbol{x},t)}{\partial \boldsymbol{x}}\cdot \frac{\partial\eta(\boldsymbol{x},t)}{\partial \boldsymbol{x}}\right]\phi_z(\boldsymbol{x},t)=0,
\label{eq:bc1}
\end{eqnarray}

\begin{eqnarray}
&&\frac{\partial\psi(\boldsymbol{x},t)}{\partial t} + \frac{1}{2}\frac{\partial\psi(\boldsymbol{x},t)}{\partial \boldsymbol{x}}\cdot\frac{\partial\psi(\boldsymbol{x},t)}{\partial \boldsymbol{x}}+\eta(\boldsymbol{x},t)\nonumber\\ 
&&-\frac{1}{2}\left[1+\frac{\partial\eta(\boldsymbol{x},t)}{\partial \boldsymbol{x}}\cdot \frac{\partial\eta(\boldsymbol{x},t)}{\partial \boldsymbol{x}}\right]\phi_z(\boldsymbol{x},t)^2=0,
\label{eq:bc2}
\end{eqnarray}
where $\phi_z(\boldsymbol{x},t)\equiv \partial \phi/\partial z|_{z=\eta}(\boldsymbol{x},t)$ is the surface vertical velocity with $\phi(\boldsymbol{x},z,t)$ being the velocity potential of the flow field, and $\psi(\boldsymbol{x},t)=\phi(\boldsymbol{x},\eta,t)$. The key procedure in HOS is to solve for $\phi_z(\boldsymbol{x},t)$ given $\psi(\boldsymbol{x},t)$ and $\eta(\boldsymbol{x},t)$, formulated as a boundary value problem for $\phi(\boldsymbol{x},z,t)$. This is achieved through a pseudo-spectral method in combination with a mode-coupling approach, with details included in multiple papers such as \cite{dommermuth1987high,pan2018high}.

By assuming small body motion, the motion of a floating ship within the time domain can be described by the Cummins equation~\citep{ogilvie1964recent} as
\begin{eqnarray} 
(\boldsymbol{M}+\boldsymbol{M}_a) \ddot {\boldsymbol{S}}(t)&+&\int_0^t\boldsymbol{K}(t-\tau)\dot{\boldsymbol{S}}(\tau)d\tau\nonumber\\
&+&\boldsymbol{C}^R\boldsymbol{S}(t)=\boldsymbol{f}^{exc}(t)
\label{eq:cum}
\end{eqnarray}
where $\boldsymbol{M}$ and $\boldsymbol{M}_a$ are the mass and added mass matrix of the floating structure. $\boldsymbol{K}(t)$ is the impulse response function considering the impact of the past motion. $\boldsymbol{C}^R$ is the radiation-restoring coefficient. $\boldsymbol{f}^{exc}(t)$ is the external excitation force, which only includes the hydrostatic restoring and Froude–Krylov forces in this study. Here we consider the linear Froude–Krylov force, which is calculated by directly integrating the pressure of undisturbed incident waves over the instantaneous submerged volume~\citep{jang2020effects}.

\subsection{Data Assimilation Scheme by EnKF}
\label{sec:enkf}
\noindent Equations \eqref{eq:bc1}, \eqref{eq:bc2}, and \eqref{eq:cum}  are integrated in time for each ensemble member to provide the ensemble of forecasts at $t=t_j$ (for $j\geq1$):
\begin{equation}
\boldsymbol{\Theta}_{\text{f},j}=[\theta_{\text{f},j}^{(1)},\theta_{\text{f},j}^{(2)},\cdots \theta_{\text{f},j}^{(n)},\cdots \theta_{\text{f},j}^{(N-1)},\theta_{\text{f},j}^{(N)}],
\label{eq:etaen}
\end{equation}
where

\begin{align}
    \theta_{\text{f},j}^{(n)} &= \begin{bmatrix}
           \eta_{\text{f},j}^{(n)} \\
           \psi_{\text{f},j}^{(n)} \\
           \boldsymbol{S}_{\text{f},j}^{(n)}
         \end{bmatrix}
\end{align}
is the $n_{th}$ member of the ensembles of model forecast results. The error covariance matrix of the model forecast can be computed  as:
\begin{equation}
\boldsymbol{Q}_{j}=\mathfrak{C}(\boldsymbol{\Theta}_{\text{f},j}),
\label{eq:covetaf}
\end{equation}
where the operator $\mathfrak{C}$ is defined as
\begin{equation}
    \mathfrak{C}(\boldsymbol{\Theta})=\boldsymbol{\Theta}'(\boldsymbol{\Theta}')^{\text{T}},
\end{equation}

\begin{eqnarray}
\boldsymbol{\Theta}'=\frac{1}{\sqrt{N-1}}[\theta^{(1)}-\bar{\theta},~\theta^{(2)}-\bar{\theta},~\dots \nonumber\\
\theta^{(n)}-\bar{\theta}~\dots ~\theta^{(N-1)}-\bar{\theta},~\theta^{(N)}-\bar{\theta}],
\label{eq:mean1}
\end{eqnarray}
\begin{equation}
\bar{\theta}=\frac{1}{N}\sum_{n=1}^{N} \theta^{\text(n)}.
\end{equation}
Also the error covariance matrix of the measurements can be calculated as
\begin{equation}
   {R}_{j}=\mathfrak{C}(\boldsymbol{\Theta}_{\text{m},j}).
\end{equation}
Here it should be noted the measurements may include those for wave, ship motion, or both, therefore the elements in $\theta_{\text{m},j}^{(n)}$ needed to be modified based on the specific situation, i.e., $\theta_{\text{m},j}^{(n)} =[\eta_{\text{m},j}^{(n)}, \psi_{\text{m},j}^{(n)}]^T$, $\theta_{\text{m},j}^{(n)} =\boldsymbol{S}_{\text{m},j}^{(n)}$, or $\theta_{\text{m},j}^{(n)} = [\eta_{\text{m},j}^{(n)}, \psi_{\text{m},j}^{(n)}, \boldsymbol{S}_{\text{m},j}^{(n)}]^T$

An analysis step is then performed, which combines the ensembles of model forecasts and perturbed measurements to produce the optimal analysis states~\citep{carrassi2018data}:
\begin{equation}
\mathop{\boldsymbol{\Theta}_{\text{a},j}}=\mathop{\boldsymbol{\Theta}_{\text{f},j}}+\mathop{\boldsymbol{K}_{j}}[\mathop{\boldsymbol{\Theta}_{\text{m},j}}-\mathop{\boldsymbol{G}_j}\mathop{\boldsymbol{\Theta}_{\text{f},j}}],
\label{eq:ana1}
\end{equation}
where 
\begin{equation}
  {\boldsymbol{K}_{j}}={\boldsymbol{Q}_{j}}{\boldsymbol{G}_j^\text{T}}\big[\boldsymbol{G}_j\boldsymbol{Q}_{j}\boldsymbol{G}_j^\text{T}+\boldsymbol{R}_{j}]^{-1} 
\label{eq:k1}
\end{equation}
is the optimal Kalman gain matrix. $\boldsymbol{G}_j$ is a linear operator, which maps a state vector from the model space to the measurement space.

Finally, to realize the simultaneous estimation or correction of parameters (e.g. ship added mass), the only action needed is to form a parameter-augmented state space by adding parameters to the original state space for the forecast and analysis, i.e.

\begin{align}
    \theta_{*,j}^{(n)} &= \begin{bmatrix}
           \eta_{*,j}^{(n)} \\
           \psi_{*,j}^{(n)} \\
           \boldsymbol{S}_{*,j}^{(n)}\\
           \boldsymbol{P}_{*,j}^{(n)}
         \end{bmatrix}
\label{eq:sp}
\end{align}
where $\boldsymbol{P}$ represents the parameters, and $*=\text{f}~\text{or}~\text{a}$.

\section{Numerical Results}
\label{sec:nr}
\begin{figure}[H]
\includegraphics[width=\linewidth]{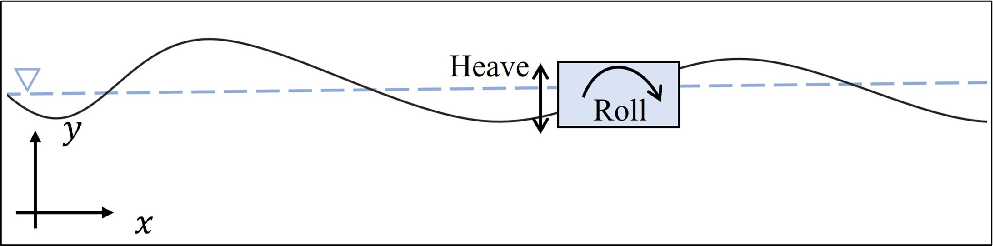}
\caption{Schematic illustration of the synthetic case.} 
\label{fig:wavefloat}
\end{figure}

To test the performance of the EnKF-HOS-CMI algorithm, we conduct a series of numerical experiments based on a synthetic case involving a 2D wave field and a two-degrees-of-freedom (i.e., heave $S_3$ and roll $S_4$) free-floating ship (fig.~\ref{fig:wavefloat}). We first use a reference HOS-CMI coupled simulation to generate the true wave field ($\eta^{\text{true}}(\boldsymbol{x},t)$ and~$\psi^{\text{true}}(\boldsymbol{x},t)$) and structure motions ($S_3^{\text{true}}(t)$ and $S_4^{\text{true}}(t)$). Specifically, we use a reference initial wave field described by a JONSWAP spectrum with a peak wavenumber $k_p=16k_0$ (with $k_0$ the fundamental wavenumber), a global steepness $k_pH_s/2=0.11$ (with $H_s$ the significant wave height) and an enhancement factor $\gamma=3.3$. $L=256$ grid points are used in spatial domain $[0,2\pi)$ in all the simulations. For the ship, the presumed reference (exact) properties
are tabulated in Table~\ref{tab:float}. We also assume that the ship is stationary initially. Then, we superpose independent random errors onto $\eta^{\text{true}}(\boldsymbol{x},t)$, $S_3^{\text{true}}(t)$, and $S_4^{\text{true}}(t)$ to produce $\eta_{\text{m},j}(\boldsymbol{x})$, $S_3^{\text{m}}(t)$, and $S_4^{\text{m}}(t)$, respectively. Finally, $\psi_{\text{m},j}(\boldsymbol{x})$ is constructed from $\eta_{\text{m},j}(\boldsymbol{x})$ based on the linear wave theory. (we refer the readers to see the detailed procedures in~\cite{wang2021phase, wang2022phase}). 

\begin{table}[H]
\caption{Properties of the ship}
\begin{small}
\begin{center}
\begin{tabular}{|c|c|c|}
\hline
Property & Symbol & Value\\
\hline
Water Plane Area & $S_w$ & $0.08\lambda_p$\\
Draft & $D$ & $2.68H_s$\\
Horizontal coordinate of centroid & $x_c$ & $1.2\pi$\\

Center of buoyancy & ${KC}$ & $-1.34H_s$\\

Center of mass & $\boldsymbol{KG}$ & $0$\\

Mass & $M$ & $3.78\times 10^{-3}$\\
Added Mass & $M_a$ & $1.31\times 10^{-3}$\\
Inertia & $I$ & $2.02\times 10^{-6}$\\
Added Inertia & $I_a$ & $9.89\times 10^{-7}$\\

\hline
\end{tabular}
\end{center}
\end{small}
\label{tab:float}
\end{table} 
To evaluate the performance of the developed EnKF-HOS-CMI algorithm, we define an error metric

\begin{equation}
\epsilon(t)=\frac{\int\mid\eta^\text{true}(\boldsymbol{x},t)-\eta^{\text{sim}}(\boldsymbol{x},t)\mid^2 dA}{2\sigma_{\eta}^2 A},
\label{eq:epsilon}
\end{equation}
where $A$ is the simulation domain, $\eta^{\text{sim}}(x,t)$ represents the simulation results obtained from EnKF-HOS-CMI (the ensemble average in this case) or the HOS-CMI (i.e. without data assimilation), and $\sigma_\eta$ is the standard deviation of $\eta^{\text{true}}$.

\begin{figure}[H]
\includegraphics[trim=10cm 0cm 10cm 0cm, clip, width=\linewidth]{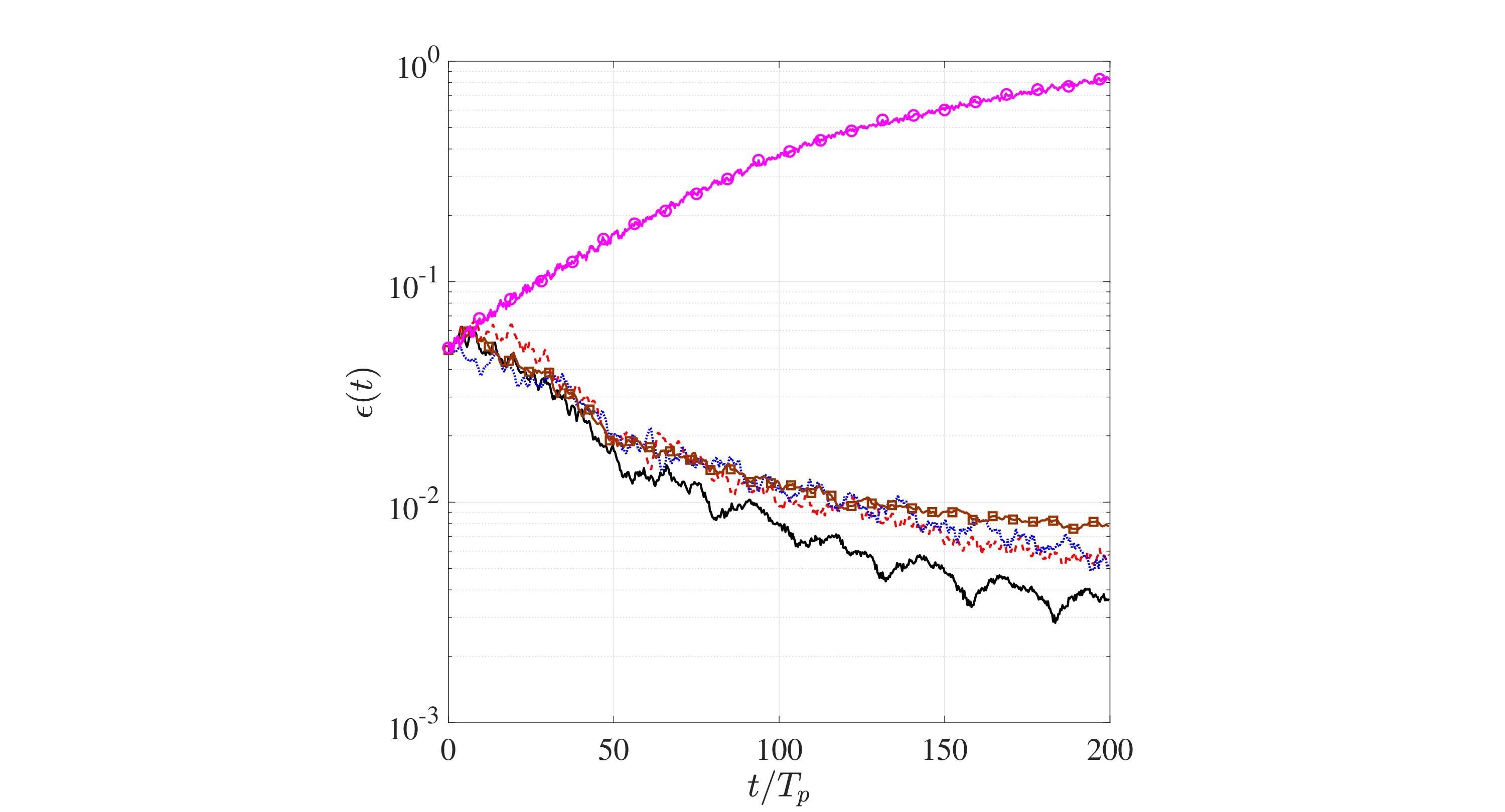}
\caption{Wave forecast errors with HOS-CMI ({\color{magenta}\Lcirc}) and EnKF-HOS-CMI by assimilating wave data ({\color{red}\dashL}); heave data ({\color{blue}\dotL}); roll data ({\color{brown}\Lbox}); wave, heave, and roll data (\rule[0.5ex]{0.5cm}{0.25pt})} 
\label{fig:err}
\end{figure}

The errors $\epsilon(t)$ obtained from HOS-CMI and EnKF-HOS-CMI are shown in figure~\ref{fig:err}. For HOS-CMI method, i.e., without DA, $\epsilon(t)$ increases in time from the initial value $\epsilon(0)\approx 0.05$, and reaches $O(1)$ at $t/T_p\approx 200$. For EnKF-HOS-CMI method, we have tested its performance by assimilating different types of data, including wave data ; heave data; roll data; combination of wave, heave, and roll data. For all cases with EnKF-HOS-CMI method, the data assimilation interval is set to be $\tau=\frac{1}{16}T_0$, where $T_0$ is the period corresponding to $k_0$. In addition, we assume that the wave data only contain those at $x=\pi$, i.e, $\eta_{\text{m},j}(x=\pi)$ and $\psi_{\text{m},j}(x=\pi)$. It can be observed that $\epsilon(t)$ from the EnKF-HOS-CMI method keeps decreasing when assimilating any type of data, and becomes three orders of magnitude smaller than that from HOS-CMI method at the end of the simulation. Also, it should be noted that when using single data (wave, heave, or roll), the errors from EnKF-HOS-CMI method are quite close to each other. While in contrast, when all the wave, heave and roll data are used, the performance of EnKF-HOS-CMI is further improved.

\begin{figure}[H]
\includegraphics[trim=10cm 0cm 10cm 0cm, clip, width=\linewidth]{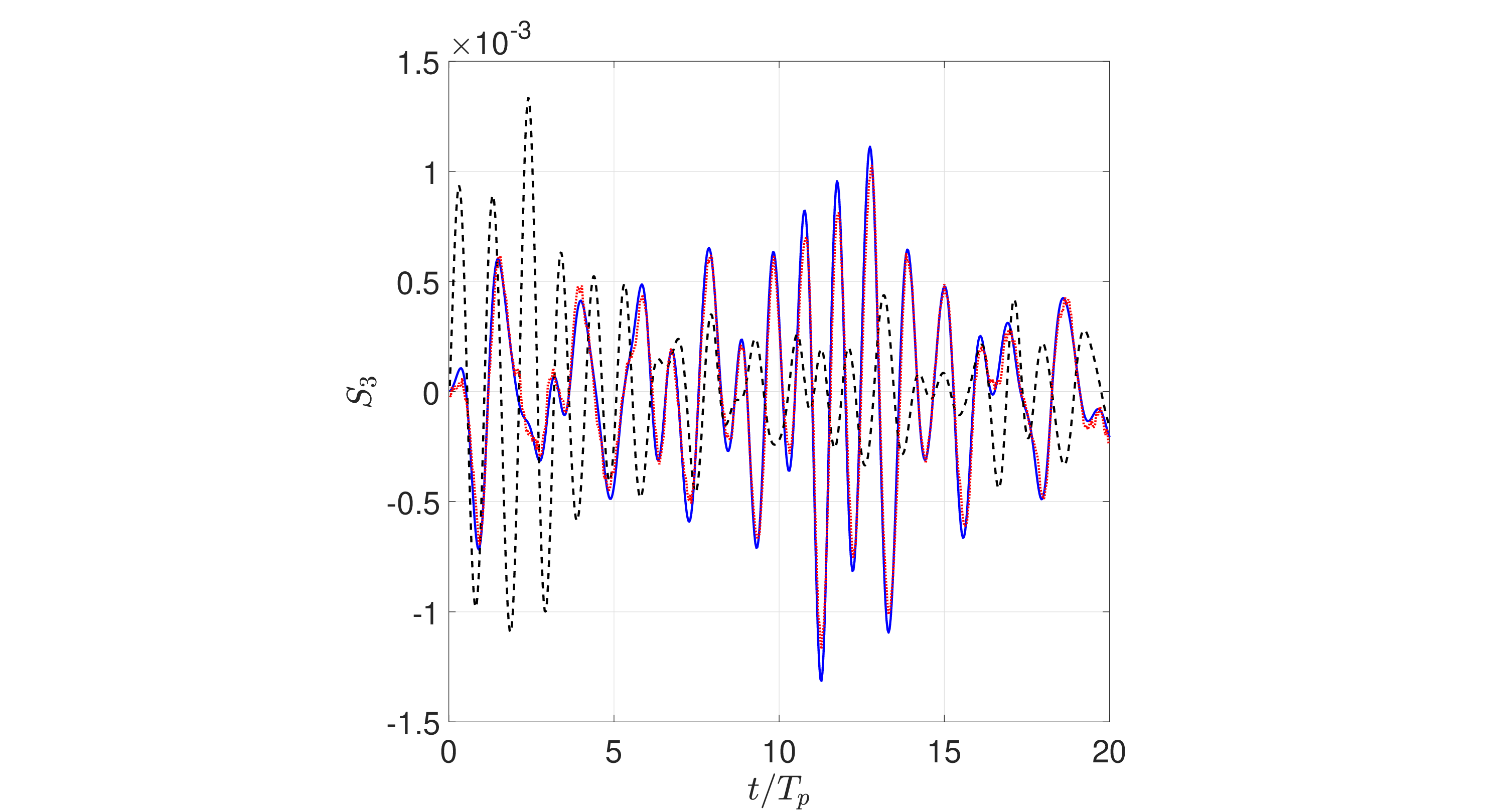}
\caption{Time histories of heave motion: reference solution ({\color{blue}\rule[0.5ex]{0.5cm}{0.25pt}}) ; w/ DA ({\color{red}\dotL}); w/o DA ({\dashL}).} 
\label{fig:heave}
\end{figure}

\begin{figure}[H]
\includegraphics[trim=10cm 0cm 10cm 0cm, clip, width=\linewidth]{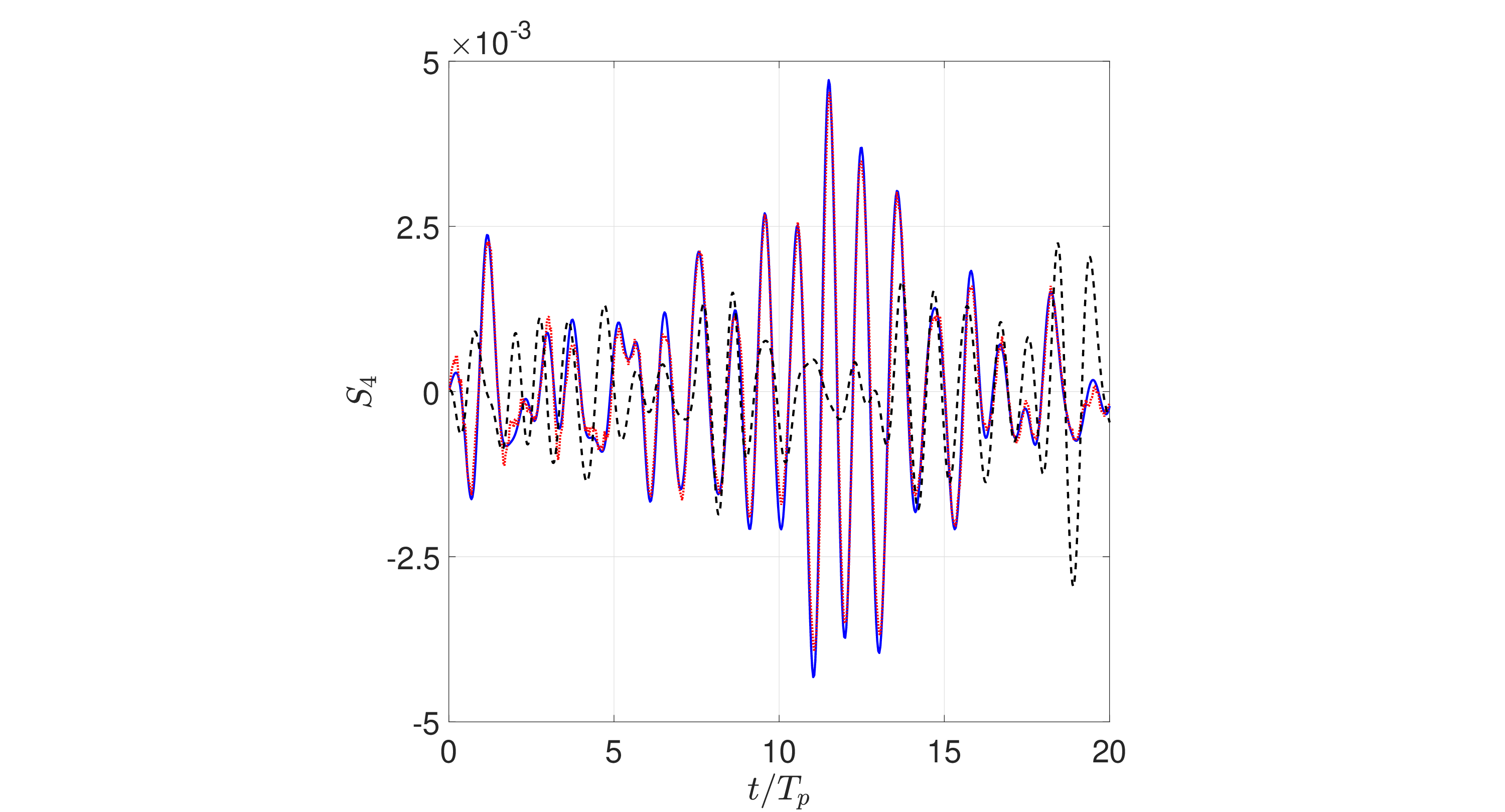}
\caption{Time histories of roll motion: reference solution ({\color{blue}\rule[0.5ex]{0.5cm}{0.25pt}}) ; EnKF-HOS-CMI ({\color{red}\dotL}); HOS-CMI ({\dashL}).} 
\label{fig:roll}
\end{figure}

\begin{figure}[H]
\includegraphics[trim=10cm 0cm 10cm 0cm, clip, width=\linewidth]{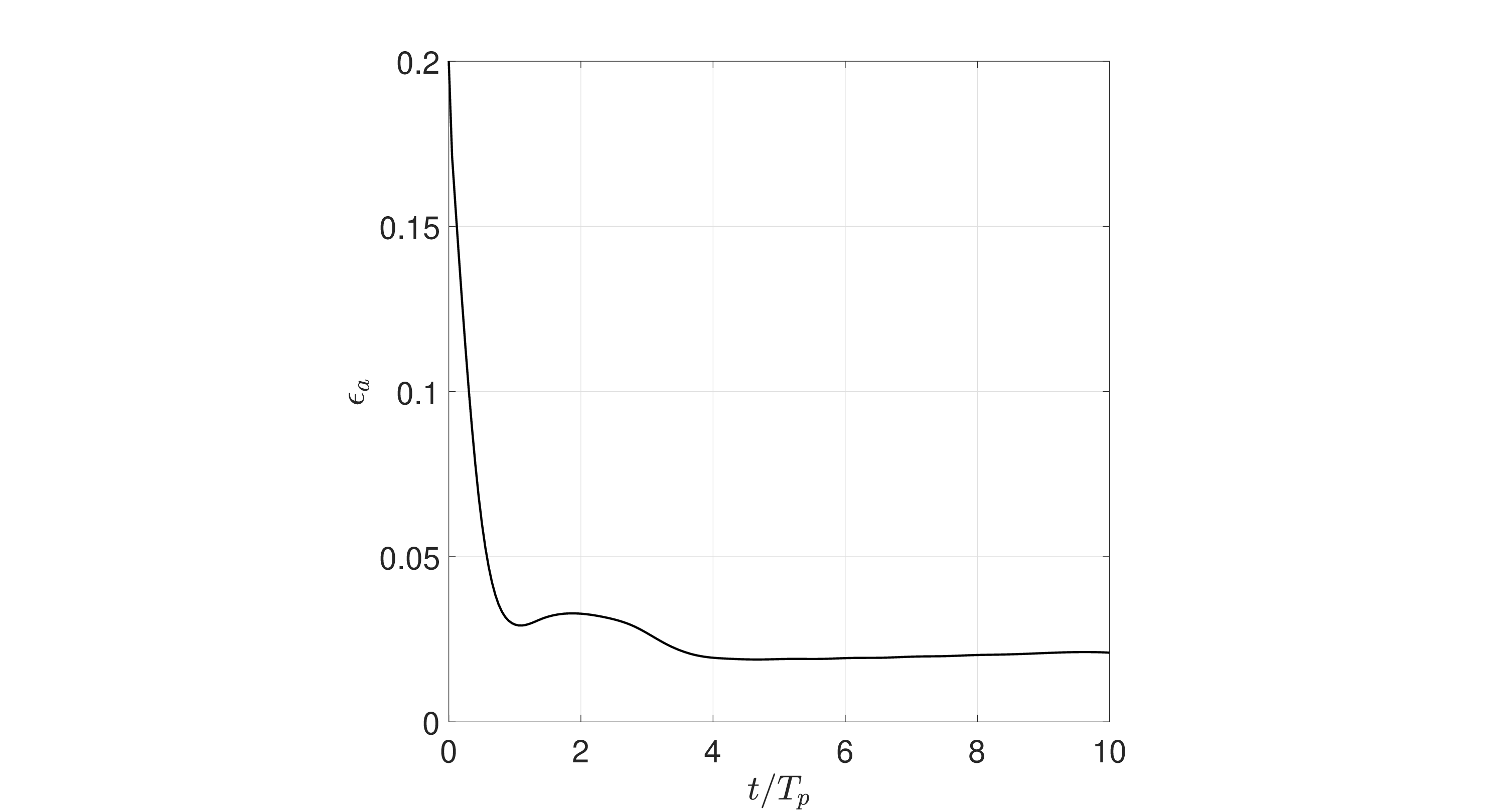}
\caption{Time history of the error for the estimated added mass.} 
\label{fig:ma}
\end{figure}

\begin{figure}[H]
\includegraphics[trim=10cm 0cm 10cm 0cm, clip, width=\linewidth]{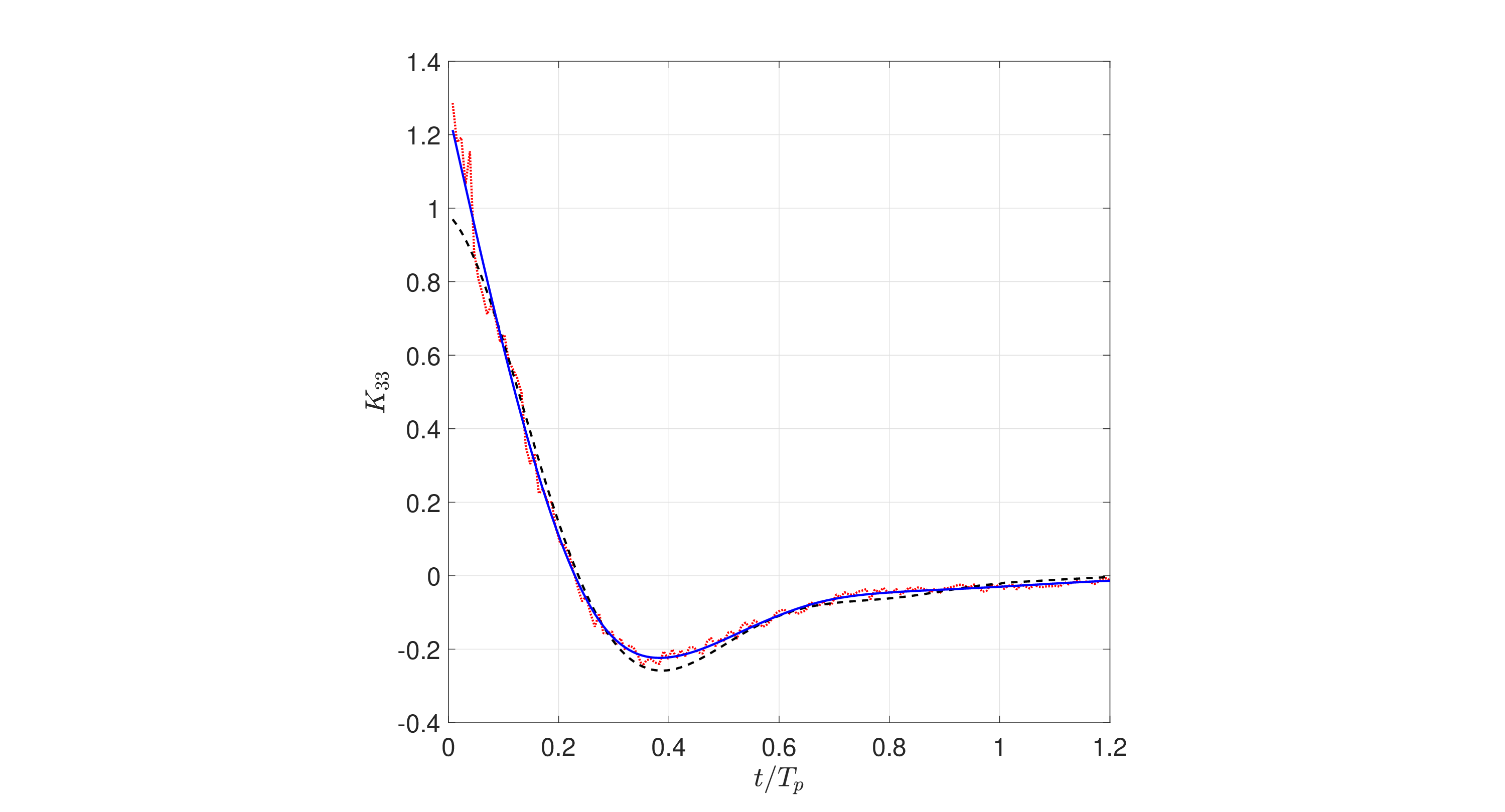}
\caption{Plots of $K_{33}$: reference ({\color{blue}\rule[0.5ex]{0.5cm}{0.25pt}}); initial guess ({\dashL}); estimated result after $t/T_p=40$ ({\color{red}\dotL}).} 
\label{fig:kt}
\end{figure}

Next, we investigate the performance of the developed EnKF-HOS-CMI method in terms of predicting ship motions. Figures~\ref{fig:heave} and~\ref{fig:roll} show the results of predicted heave and roll motions, respectively, obtained with EnKF-HOS-CMI and HOS-CMI simulations, in comparison with the reference solutions. For both EnKF-HOS-CMI and HOS-CMI simulations, the initial states of the floating ship are assumed to be stationary, which is the same as the reference solution. In addition, we assume the added mass ($M_a$) and impulse response function $\boldsymbol{K}(t)$ have initial deviations from the truth (as shown Figures~\ref{fig:ma} and~\ref{fig:kt}), and the simultaneous estimation (correction) of them is be realized by setting $\boldsymbol{P}^{(n)}_{*,j}$ in Equation~\eqref{eq:sp} to be
\begin{equation}
    P^{(n)}_{*,j}=[M_{a,*,j}^{(n)},\boldsymbol{K}_{*,j}^{(n)}]^T
\end{equation}
It can be found that even if the initial errors for the ship motion is set to be zero, the predicted heave and roll motions by the HOS-CMI method quickly deviate from the truth, which is due to mis-estimation of the wave filed, i.e. excitation force, and inaccurate structural parameters. In contrast, when using EnKF-HOS-CMI, even though some (mild) deviations from the truth can be observed (especially for the local extreme points), the results can always follow the overall trend of the reference solutions. Figure~\ref{fig:ma} shows time history of the error of the estimated added mass.  It can be found that, as the observed data are assimilated, the error of estimated added mass quickly decreases (although experiencing some fluctuations), and after only $t/T_p=4$, the error is already reduced by 10 times in comparison with its initial value. Figure~\ref{fig:kt} shows the estimated impulse response function with the EnKF-HOS-CMI after $t/T_p=40$, and its deviation from the reference is already negligible. 

\section{Conclusion}
\label{SECconclusion}

In this paper, we develop an EnKF-HOS-CMI coupled algorithm,  which is aimed at further improvement of the robustness of the DA-based phase-resolved ocean wave forecast and realizing simultaneous ship motion prediction. The developed algorithm is tested and validated based on the coupled simulation of an irregular wave field and a boxed-shaped free-floating ship. It is shown that, by assimilating ship motion data (either heave or roll) into the HOS-CMI model, the wave forecast accuracy is quite close to that given by assimilating the wave data. Also, it has been demonstrated that the wave forecast accuracy can be further improved by using both the wave and ship motion data. We have also tested the performance of the EnKF-HOS-CMI method in terms of ship motion prediction and estimating ship parameters. It is indicated this method can not only accurately predict the ship motion, but also realize the high-accuracy estimation of parameters. In the future, the developed EnKF-HOS-CMI algorithm will be extended to three-dimensional problems, and real wave and ship data will be used to benchmark and calibrate the method.

%\section{Acknowledgements}
%\label{SECacknowledgements}

%This work was supported by ...
\bibliographystyle{plainnat}
\bibliography{asme2e}

\begin{thebibliography}{15}
\providecommand{\natexlab}[1]{#1}
\providecommand{\url}[1]{\texttt{#1}}
\expandafter\ifx\csname urlstyle\endcsname\relax
  \providecommand{\doi}[1]{doi: #1}\else
  \providecommand{\doi}{doi: \begingroup \urlstyle{rm}\Url}\fi

\bibitem[Carrassi et~al.(2018)Carrassi, Bocquet, Bertino, and Evensen]{carrassi2018data}
Alberto Carrassi, Marc Bocquet, Laurent Bertino, and Geir Evensen.
\newblock Data assimilation in the geosciences: An overview of methods, issues, and perspectives.
\newblock \emph{Wiley Interdisciplinary Reviews: Climate Change}, 9\penalty0 (5):\penalty0 e535, 2018.

\bibitem[Chen et~al.(2020)Chen, Okada, Kawamura, and Mitsuyuki]{chen2020estimation}
Xi~Chen, Tetsuo Okada, Yasumi Kawamura, and Taiga Mitsuyuki.
\newblock Estimation of on-site directional wave spectra using measured hull stresses on 14,000 teu large container ships.
\newblock \emph{Journal of Marine Science and Technology}, 25:\penalty0 690--706, 2020.

\bibitem[Dommermuth and Yue(1987)]{dommermuth1987high}
Douglas~G Dommermuth and Dick~KP Yue.
\newblock A high-order spectral method for the study of nonlinear gravity waves.
\newblock \emph{Journal of Fluid Mechanics}, 184:\penalty0 267--288, 1987.

\bibitem[Jang and Kim(2020)]{jang2020effects}
HaKun Jang and MooHyun Kim.
\newblock Effects of nonlinear fk (froude-krylov) and hydrostatic restoring forces on arctic-spar motions in waves.
\newblock \emph{International Journal of Naval Architecture and Ocean Engineering}, 12:\penalty0 297--313, 2020.

\bibitem[Koterayama et~al.(2002)Koterayama, Nakamura, Ikebuchi, Takatsu, Fujii, and Sato]{koterayama2002study}
Wataru Koterayama, Masahiko Nakamura, Tetsuro Ikebuchi, Naoyuki Takatsu, Satoshi Fujii, and Kenji Sato.
\newblock A study for development of a wave observation platform.
\newblock \emph{Journal of the Society of Naval Architects of Japan}, 2002\penalty0 (191):\penalty0 57--67, 2002.

\bibitem[Nielsen and Dietz(2020)]{nielsen2020ocean}
Ulrik~D Nielsen and Jesper Dietz.
\newblock Ocean wave spectrum estimation using measured vessel motions from an in-service container ship.
\newblock \emph{Marine Structures}, 69:\penalty0 102682, 2020.

\bibitem[Nielsen(2006)]{nielsen2006estimations}
Ulrik~Dam Nielsen.
\newblock Estimations of on-site directional wave spectra from measured ship responses.
\newblock \emph{Marine Structures}, 19\penalty0 (1):\penalty0 33--69, 2006.

\bibitem[Ogilvie(1964)]{ogilvie1964recent}
T~Francis Ogilvie.
\newblock Recent progress toward the understanding and prediction of ship motions.
\newblock In \emph{David W. Taylor Model Basin, Washington DC, USA, Presented at: Proceedings of the 5th Symposium on Naval Hydrodynamics, Bergen, Norway, pp. 3-80}, 1964.

\bibitem[Pan et~al.(2018)Pan, Liu, and Yue]{pan2018high}
Yulin Pan, Yuming Liu, and Dick~KP Yue.
\newblock On high-order perturbation expansion for the study of long--short wave interactions.
\newblock \emph{Journal of Fluid Mechanics}, 846:\penalty0 902--915, 2018.

\bibitem[Pascoal et~al.(2005)Pascoal, Guedes~Soares, and Sorensen]{pascoal2005ocean}
R~Pascoal, C~Guedes~Soares, and AJ~Sorensen.
\newblock Ocean wave spectral estimation using vessel wave frequency motions.
\newblock In \emph{International Conference on Offshore Mechanics and Arctic Engineering}, volume 41960, pages 337--345, 2005.

\bibitem[Takami et~al.(2022)Takami, Nielsen, and Jensen]{takami2022application}
Tomoki Takami, Ulrik~Dam Nielsen, and J{\o}rgen~Juncher Jensen.
\newblock Application of prolate spheroidal wave functions for assessment and prediction of ship responses.
\newblock In \emph{15th International Symposium on Practical Design of Ships and Other Floating Structures}, 2022.

\bibitem[Takami et~al.(2023)Takami, Nielsen, Jensen, and Chen]{takami2023estimation}
Tomoki Takami, Ulrik~Dam Nielsen, J{\o}rgen~Juncher Jensen, and Xi~Chen.
\newblock Estimation of encountered wave elevation sequences based on response measurements in multi-directional seas.
\newblock \emph{Applied Ocean Research}, 135:\penalty0 103570, 2023.

\bibitem[Tannuri et~al.(2003)Tannuri, Sparano, Simos, and Da~Cruz]{tannuri2003estimating}
Eduardo~A Tannuri, Joao~V Sparano, Alexandre~N Simos, and Jos{\'e}~J Da~Cruz.
\newblock Estimating directional wave spectrum based on stationary ship motion measurements.
\newblock \emph{Applied Ocean Research}, 25\penalty0 (5):\penalty0 243--261, 2003.

\bibitem[Wang and Pan(2021)]{wang2021phase}
Guangyao Wang and Yulin Pan.
\newblock Phase-resolved ocean wave forecast with ensemble-based data assimilation.
\newblock \emph{Journal of Fluid Mechanics}, 918:\penalty0 A19, 2021.

\bibitem[Wang et~al.(2022)Wang, Zhang, Ma, Zhang, Li, and Pan]{wang2022phase}
Guangyao Wang, Jinfeng Zhang, Yuxiang Ma, Qinghe Zhang, Zhilin Li, and Yulin Pan.
\newblock Phase-resolved ocean wave forecast with simultaneous current estimation through data assimilation.
\newblock \emph{Journal of Fluid Mechanics}, 949:\penalty0 A31, 2022.

\end{thebibliography}

\end{multicols*}
\end{document}